# The electrification of granular systems of identical insulators

**Jasper F. Kok[1,2,*] and Daniel J. Lacks[3]**
[1]Applied Physics Program, University of Michigan, Ann Arbor, MI, 48109, USA.
[2]Atmospheric, Oceanic, and Space Sciences, University of Michigan, Ann Arbor, MI, USA.
[3]Department of Chemical Engineering, Case Western Reserve University, Cleveland, OH 44106, USA

**ABSTRACT**

Insulating particles can become highly electrified during powder handling, volcanic eruptions, and the wind-blown transport of dust, sand, and snow. Measurements in these granular systems have found that smaller particles generally charge negatively, while larger particles charge positively. These observations are puzzling, since particles in these systems are generally chemically identical, and thus have no contact potential difference. We show here that simple geometry leads to a net transfer of electrons from larger to smaller particles, in agreement with these observations. We integrate this charging mechanism into the first quantitative charging scheme for a granular system of identical insulators, and show that its predictions are in agreement with measurements. Our theory thus seems to provide an explanation for the hitherto puzzling phenomenon of the size-dependent charging of granular systems of identical insulators.

# I. INTRODUCTION

It has been known since ancient times that two objects rubbed together can charge each other [1,2]. This frictional or 'triboelectric' charge transfer also occurs between two insulators, which is quite remarkable considering that insulators do not contain free charge carriers [2-4]. Even more remarkable is the observation that triboelectric charging also occurs between insulating particles that are chemically identical (i.e., that have the same contact potential) [4-8]. For granular systems, the charging is generally such that smaller particles charge negative while larger particles charge positive. This size-dependent charging of identical insulators has recently been demonstrated in fluidized beds [9], and is suggested by measurements in a wide variety of other granular systems, including powder handling [10], wind-blown sand on Earth [11,12] and possibly Mars [13], wind-blown snow [14], and volcanic eruptions [15]. Determining the physical mechanism behind this perplexing phenomenon could produce fundamental advances in electrophotography technology [16], and further our understanding of the electrification of sand, dust, ash, powder, snow, and ice, and the subsequent occurrence of discharges and associated chemical reactions in these systems [8,10,17-21].

Several physical models have been proposed to explain the triboelectric charging of identical insulators. Henry [6] conjectured that the charge transfer was due to temperature gradients caused by asymmetry in the rubbing, but experiments by Lowell and Truscott [7,8] showed that the charging does not depend on the rubbing speed, which affects the temperature gradient, but only on the total distance over which one object is rubbed over the other. Based on these experimental results, Lowell and Truscott [8] proposed that the charge transfer is instead due to the presence of high-energy electrons 'trapped' in defect states [22]. These states cannot equilibrate with nearby empty low-energy states on the same particle because of energetic constraints [2,8]. The existence of these trapped high-energy electron states is supported by the occurrence of phosphorescence and thermoluminescence phenomena in insulators [23-25]. However, if surface contact brings empty low-energy states on another particle within close enough proximity, trapped high-energy electrons could tunnel to those states [8]. An asymmetry in the nature of the surface contact between the two objects could thus produce an imbalance in the number of transferred electrons and therefore produce a net transfer of charge. For example, the experiments of Shaw, Henry, and Lowell and Truscott [5-8] showed that rubbing a small area of one object over a large area of another, identical, object causes the former object to obtain a negative charge.

For granular systems of chemically-identical insulators, Lacks et al. [26] recently showed that a systematic charge transfer is produced by a different asymmetry. They showed that, after several initial collisions in which small and large colliding particles lose roughly equal amounts of trapped electrons, smaller particles have nonetheless lost a larger *fraction* of their trapped electrons than larger particles have. Therefore, in subsequent collisions, smaller particles give up fewer trapped electrons than larger particles do, leading smaller particles to charge negatively and larger particles to charge positively, in agreement with the measurements discussed above [9-12,14,15].

We here identify a second mechanism that causes small particles to charge negative and large particles to charge positive. We show that simple geometric considerations cause more electrons to tunnel from the larger particle to the smaller particle than vice versa. We combine this charging mechanism with the 'multiple collisions' mechanism discussed above, and develop the first quantitative charging scheme for a granular system of identical insulators. The scheme also accounts for the effect of particle charge on subsequent charge transfer, following ideas inherent in the surface state theory [27-30], which describes charge transfer between dissimilar insulators and between metals and insulators. The predictions of our charging scheme are in both qualitative and quantitative agreement with measurements.

In the next section we describe our theoretical model and discuss its assumptions. In Section III, we then use this model to derive the geometric charge transfer due to differences in particle size, formulate a quantitative charging scheme, and compare the predictions of this charging scheme to measurements. In Section IV, we discuss the limitations of our charging scheme, its relation to the surface state theory, and a semi-empirical adaptation of our model.

## II. THEORETICAL MODEL

We consider a granular system of idealized spherical insulating particles that are chemically identical. For simplicity, we assume that electrons on the surface of these particles can be in either a low-energy (L) or a high-energy (H) state [8,26]. The number of high and low-energy electrons on the particle's surface at time $t$ are denoted as $n_{iH}(t)$ and $n_{iL}(t)$, and the initial number of high-energy electrons is given by

$$n_{iH}(0) = 4\pi R_i^2 \rho_{H,0}, \tag{1}$$

where $R_i$ is the radius of particle $i$. We assume the initial density of high-energy surface states, $\rho_{H,0}$, to be equal for all particles [26,31]. The surface density of low-energy electrons is probably several orders of magnitude larger [8], but following previous investigators [8,26,31] we assume that their contribution to the charge transfer is negligible, and the initial value $n_{iL}(0)$ is thus irrelevant.

We assume that, during a collision, electrons can relax from high-energy states on one particle to low-energy states on the opposite particle. Specifically, we assume that all high-energy electrons within a distance $\delta_0$ of the surface of the opposite particle (Fig. 1a) will tunnel to empty low-energy states on that particle's surface [8,26,31]. Lowell [32] modeled this electron transfer process in terms of the tunneling dynamics of a particle in a one dimensional square well separated by an energy barrier from another square well (Fig. 1b). He showed that the maximum distance $\delta_0$ that an electron in the ground state of a square well can tunnel during a collision is approximately given by

$$\delta_0 = \frac{\hbar}{\sqrt{8mE_b}} \ln\left(\frac{\hbar}{\eta m} \frac{t_{coll}}{a^2}\right), \tag{2}$$

where $\eta = (1.12/\pi^2)(2+\pi)$, $E_b$ is the height of the energy barrier between the two potential wells (see Fig. 1b), $m$ is the electron mass, $\hbar$ is the reduced Planck constant, $t_{coll}$ is the time scale of the collision, and $a$ is the radius of the well corresponding to the electron trap. Since electrons can transfer between particles during collisions, net charges can develop on the particles. The net charge on a particle of type $i$ is then given by

$$q_i(t) = e[n_{iH}(0) - n_{iH}(t) + n_{iL}(0) - n_{iL}(t)], \tag{3}$$

where $e$ is the elementary charge.

If the colliding particles hold net charges, then the electrostatic potential difference $\Delta V$ between their surfaces will make the transfer of electrons from the negatively charged to the positively charged particle more energetically favorable than vice versa [27,30]. Indeed, this effect will alter the barrier height such that

$$E_b = E_{b0} + e\Delta V, \quad (4)$$

where the potential difference $\Delta V$ between colliding particles with charges $q_i$ and $q_j$ is approximately given by [2,30]

$$\Delta V = \frac{\delta_0}{4\pi\varepsilon_0}\left(\frac{q_i}{R_i^2} - \frac{q_j}{R_j^2}\right). \quad (5)$$

Note that $\Delta V$ and $\delta_0$ are interdependent, such that Eqs. (2) and (5) must be solved iteratively.

## III. RESULTS

### A. Geometric effect in a single collision

We use the theoretical model described above to study the charge transfer between colliding idealized spherical particles of different sizes due to the tunneling of trapped high-energy electrons to empty low-energy states on the opposite particle. We hypothesize that this transfer of electrons is proportional to the particle's surface area that, at the instant of collision, is within the distance $\delta_0$ of the surface of the opposite particle (see Fig. 1a).

From Figure 1a, the number of electrons transferred from the high energy states of particle $i$ ($-\Delta n_{iH}$) to the empty low energy states of particle $j$ ($\Delta n_{jL}$), is equal to

$$-\Delta n_{iH} = \Delta n_{jL} = 2\pi\rho_{H,0}R_i^2(1-\cos\theta_i). \quad (6)$$

The angle $\theta_i$ represents the maximum angle from the contact point for which the surface of particle $i$ is within the distance $\delta_0$ of the surface of particle $j$ (see Fig. 1a), and satisfies

$$R_i \sin\theta_i = (R_j + \delta_0)\sin\alpha_j; \quad (7a)$$
$$R_i \cos\theta_i + (R_j + \delta_0)\cos\alpha_j = R_i + R_j, \quad (7b)$$

where the angle $\alpha_j$ is defined in Fig. 1a. Solving (7a) for $\alpha_j$, using that $\cos(\arcsin x) = \sqrt{1-x^2}$, squaring both sides, and solving for $\cos\theta_i$, we obtain

$$\cos\theta_i = 1 - \frac{\delta_0}{2R_i}\frac{2R_j + \delta_0}{R_i + R_j}. \quad (8)$$

Substituting this result into (6) then yields the number of electrons transferred between the colliding particles in terms of the particle sizes, the density of trapped states, and the tunneling distance

$$-\Delta n_{iH} = \Delta n_{jL} = \pi\rho_{H,0}R_i\delta_0\frac{2R_j + \delta_0}{R_i + R_j}. \quad (9)$$

By substituting (9) into (3) we then obtain the net charge transfer experienced by particle $i$ as

$$\Delta q_i = \pi e \rho_{H,0} \delta_0^2 \left( \frac{R_i - R_j}{R_i + R_j} \right). \tag{10}$$

Simple geometry thus leads the larger colliding particle to obtain a positive charge denoted by (10), while the smaller particle loses a charge of the same magnitude. This net transfer of charge increases with the imbalance in particle sizes, in agreement with experimental observations [9,12,33,34].

Although the above result – that in a single collision the larger particle will charge positively and the smaller particle will charge negatively – is valid for idealized spherical particles only, we show in the Appendix that a similar charge transfer occurs during collisions between cubical particles. Since most natural particles can probably be described as a superposition of spherical and cubical shapes, we argue that the size-dependent charging of spherical and cubical particles can be generalized to natural particles.

### B. Charge transfer scheme including multiple collisions

In addition to the geometric charge transfer mechanism identified above, identical insulators can also charge through undergoing multiple collisions [26,31]. In essence, these two charging mechanisms are different manifestations of the same physical process: the transfer of trapped high-energy electrons during collisions [8]. In this section, we thus seek to derive a charging scheme that unifies these two related charging mechanisms.

Consider the granular system of identical insulators described in Section II. The change in the number of low-energy and high-energy electrons due to charge transfer during particle collisions is given by [26]

$$\frac{dn_{iH}(t)}{dt} = -\sum_j N_j \sigma_{ij} \overline{v_{rel}^{ij}} C^{ij} n_{iH}(t), \text{ and} \tag{11a}$$

$$\frac{dn_{iL}(t)}{dt} = \sum_j N_j \sigma_{ij} \overline{v_{rel}^{ij}} C^{ji} n_{jH}(t), \tag{11b}$$

where $j$ sums over all particle sizes present in the system, $N$ denotes the particle density, and $\sigma_{ij} = \pi(R_i + R_j)^2$ and $\overline{v_{rel}^{ij}}$ are respectively the collisional cross section and the average relative velocity between particles $i$ and $j$. The chance $C^{ij}$ that a given high-energy electron on particle $i$ will transfer to an empty low-energy state on particle $j$ depends on the fraction of the particle's surface area that is close enough to the opposite particle's surface (Fig. 1a) to allow high-energy electrons to tunnel across. We determine this fraction from Eq. (9), obtaining

$$C^{ij} = \frac{\delta_0}{4\pi R_i} \frac{2R_j + \delta_0}{R_i + R_j}. \tag{12}$$

The time-evolution of charges on particles in a granular system can thus be obtained by numerically solving Eq. (11), and using Eqs. (2–5,12) to obtain $C^{ij}$.

In addition to this numerical solution, we can obtain an analytical solution in the limit where the potential difference between oppositely charged particles is not large enough to significantly affect the charge transfer (that is, $e\Delta V \ll E_{b0}$). In this case, we can substitute (12) into (11a) and solve for $n_{iH}$, yielding

$$n_{iH}(t) = n_{iH}(0)\exp(-t/\tau_i), \tag{13}$$

where the decay time constant $\tau_i$ is given by

$$\tau_i = \frac{4R_i}{\pi\delta_0 \sum_j N_j \overline{v_{rel}^{ij}}(R_i + R_j)(2R_j + \delta_0)}. \tag{14}$$

Using Eqs. (12–14) to solve (11b) for $n_{iL}$ then yields

$$n_{iL}(t) = \pi^2 \rho_{H,0} \sum_j N_j \overline{v_{rel}^{ij}} R_j \delta_0 (R_i + R_j)(2R_i + \delta_0)\tau_j \left[1 - \exp(-t/\tau_j)\right] + n_{iL}(0), \tag{15}$$

such that the net charge over time on a particle is given by

$$q_i(t) = 4\pi R_i^2 e \rho_{H,0}\left[1 - \exp(-t/\tau_i)\right]$$
$$- \pi^2 e \rho_{H,0} \sum_j N_j \overline{v_{rel}^{ij}} R_j \delta_0 (R_i + R_j)(2R_i + \delta_0)\tau_j \left[1 - \exp(-t/\tau_j)\right]. \tag{16}$$

Eqs. (11) and (16) are respectively the first quantitative numerical and analytical expressions of the triboelectric charging of a granular system of identical insulators.

## C. Quantitative application of our charging scheme

To apply the charging scheme developed in the previous section, we must assign values for the parameters $\rho_{H,0}$, $E_{b0}$, and the ratio $t_{coll}/a^2$. We take $t_{coll}/a^2 = 1$ ns/Å$^2$, based on the estimates $a = 1$ Å [32] and $t_{coll} = 1$ ns for two particles that collide with a characteristic speed of 1 m/s and interact over a length scale of ~1 nm [2,32]. Note, however, that the results depend only very weakly on the ratio $t_{coll}/a^2$, due to the logarithm term in (2). The model is more sensitive to the values of $\rho_{H,0}$ and $E_{b0}$, and we thus present results over the range of realistic values of these parameters.

Results of the model, with the parameter values described in the previous paragraph, are shown in Figure 2. These results are obtained by the numerical solution of (11) and the analytical solution of (16). We find that the charge transfer is usually dominated by the multiple collisions mechanism [26,31], except when only a few collisions occur, in which case geometric charging (Eq. 10) dominates.

Figure 2a shows the dependence of the surface charge density on $\rho_{H,0}$ for a binary mixture of two particle sizes. The value of $\rho_{H,0}$ is limited between a lower bound of the typical charge density generated in granular systems of identical insulators (~$10^{14}$ elementary charges per m$^2$ [11,35]) and an upper bound of ~1 trapped electron per atom (~$10^{20}$ states per m$^2$). The predicted surface charge density (see Fig. 2a) over this wide range of $\rho_{H,0}$ is on the order of magnitude found in experiments [11,34,35]. Moreover, the magnitude of the surface charge density depends linearly on $\rho_{H,0}$ for small values, but is independent of $\rho_{H,0}$ for larger values of $\rho_{H,0}$, because the electrostatic potential difference $\Delta V$ between oppositely charged particles limits any further charge transfer (see Eqs. 2-5).

The surface charge density at which the transition between these "low-density" and "high-density" regimes occurs depends on the height of the energy barrier $E_{b0}$ (see Fig. 2b). A reasonable upper limit on $E_{b0}$ is the equivalent Fermi level of insulators, which is around 4.5 eV [2,32]. A detailed electronic structure analysis would be necessary to determine a more precise estimate for $E_{b0}$. Note that the transition between the low-density and the high-density regime can be very sharp (see also Fig. 3a), because of the interdependence between the electrostatic potential difference between colliding particles ($\Delta V$) and the tunneling distance ($\delta_0$). Once $\Delta V$ becomes on the order of $E_{b0}$, it substantially increases $\delta_0$ (see Fig. 3b), which in turn increases $\Delta V$ (see Eqs. 2-5). This positive feedback between $\Delta V$ and $\delta_0$ produces a sensitive dependence of the tunneling distance on the particle charges, leading to the sharp transition between the low-density and high-density regimes for large $t/\tau$ in Figures 2 and 3. This effect is probably not realistic for an ensemble-averaged particle charge, because the averaging over many particles with somewhat different charges would yield a smoother transition.

The normalized charging of a binary mixture of two particle sizes with time is shown in Figure 3a. Note that the predicted characteristic charging time of several minutes agrees well with measurements in fluidized beds [36-38].

It is noteworthy that, despite the large uncertainty in the value of $\rho_{H,0}$, our charging scheme is in quantitative agreement with measurements even without the use of empirical parameters (see Section IVc). Indeed, the agreement with measurements is optimal for $\rho_{H,0} \approx 10^{16}$ m$^{-2}$. In Figure 4, we apply this value of $\rho_{H,0}$ to the charging of dust and sand in dust devils and dust storms [20], and again find that our charging scheme predicts particle charges of the same order of magnitude as measurements [11,35].

## IV. DISCUSSION

### A. Relation to the surface state theory

The results in Figures 2 and 4 suggest that the observed size-dependent electrification of granular systems of identical insulators is due to the presence of trapped high-energy electrons. This result raises several intriguing questions, for instance whether trapped high-energy electrons could also play a role in charge transfer between dissimilar insulators and between metals and insulators. Charge transfer in these systems is generally well-described by the surface state theory [16,27-30], which predicts a low-density limit, where the final particle charge depends on the density of mobile electron states, and a high-density limit, where the final particle charge is independent of the density of states and is instead constrained by the electrostatic potential difference between particles. These limits are identical to those seen in Figures 2 and 3. This agreement is not surprising, since our theoretical model and the surface state theory share a very similar physical basis. However, our model proposes a specific mechanism for charge mobility: the presence of trapped high-energy electrons. The surface state theory, on the other hand, simply hypothesizes that mobile electron surface states exist which provide the charge transfer. It seems plausible then that the mobile electron states hypothesized in the surface state theory are in fact trapped high-energy electrons, whose existence is supported by the occurrence of phosphorescence and thermoluminescence in insulators [23-25]. The net charging observed between metals and insulators and between dissimilar insulators could then be due to material-dependent differences in the density ($\rho_{H,0}$) and energy level ($E_{b0}$) of the trapped high-energy electrons [39].

Another important question raised by our results is whether charging in granular systems of identical insulators is governed by the low-density or the high-density limit (see Fig. 2). While this question is difficult to resolve in the absence of detailed experimental studies, the parallel o the surface state theory, which correctly describes measurements only in the high-density limit [16,27], suggests that the high-density limit is the correct regime. Note that the final particle charges in the high-density limit are over two orders of magnitude larger than found by measurements in wind-blown sand and dust devils (Fig. 2a). While this is consistent with our result that the characteristic charging time scale of these phenomena is much larger than their lifetime (see Fig. 4), a potential alternative explanation could be that the high-density limit occurs at a much lower electrostatic potential between particles than our theory predicts. Indeed, in the surface state theory the particle charges in the high-density limit are approximately two orders of magnitude smaller than simple theoretical calculations indicate [16,27] and so a similar effect might occur for identical insulators. A potential cause of this effect might be the occurrence of micro-discharges between colliding particles (see the next section). Experimental studies are required to resolve these questions.

## B. Limitations of the theoretical model

Our theoretical model (see Section II) is necessarily idealized and neglects certain processes that could affect the charge transfer, especially for particles holding large surface charge densities. First, the model neglects the transfer of low-energy electrons. This assumption is probably justified for low values of the particle charge. However, when particle charges increase, the energy of low-energy electrons on one particle can be significantly higher than that of empty low-energy states on the oppositely charged particle, which would lead to tunneling of these low-energy electrons. A more detailed model should consider this effect.

A second limitation of the present model is that it does not account for the occurrence of electric discharges between oppositely charged colliding particles. Such discharges can occur if the electric field between the particles exceeds the breakdown electric field described by the 'Paschen law' [19,40],

$$E_{br} = \frac{BPT_0/T}{C + \ln(PLT_0/T)}, \qquad (17)$$

where $C = \ln[A/\ln(1/\gamma + 1)]$, $P$ and $T$ are the gas pressure and temperature, and $L$ is the distance over which the discharge occurs. The constants $A$, $B$, and $\gamma$ determine the ionization coefficients [40] at $T_0 = 293$ K for different gases. An electric discharge will thus occur if the electric field a distance $L$ from the surface of a charged particle with surface charge density $\sigma$ exceeds the breakdown field $E_{br}$. From Gauss' law, the electric field produced by the particle equals

$$E_p = \frac{\sigma R^2}{\varepsilon_0 (R+L)^2}. \qquad (18)$$

Solving for the minimum surface charge density $\sigma_{br}$ at which electric discharges occur for the terrestrial atmosphere (i.e., $P = 10^5$ Pa, $T = 288$ K, $A = 15$ m$^{-1}$Pa$^{-1}$, $B = 365$ Vm$^{-1}$Pa$^{-1}$, and $\gamma = 0.01$ [40]), we find that $\sigma_{br}$ respectively equals 0.3 and 0.09 mC/m$^2$ for particles of 10 and 100 μm diameter. Conversely, for the Martian atmosphere (i.e., $P \approx 700$ Pa, $T \approx 230$ K, $A = 15$ m$^{-1}$Pa$^{-1}$, $B = 350$ Vm$^{-1}$Pa$^{-1}$, and $\gamma = 0.01$ [19,40]), $\sigma_{br}$ respectively equals 2.4 and 0.04 mC/m$^2$ for particles of 10 and 100 μm diameter. For the 100 μm particles, $\sigma_{br}$ is thus on the order of surface charge densities predicted in Figures 2 and 4. Depending on the density of trapped high-energy electrons (Fig. 2a) and the energy barrier (Fig. 2b), the magnitude of the particle charging could thus be limited by the occurrence of 'micro-discharges' between colliding particles, especially for large values of $t/\tau$. Such discharges often occur during experiments with dissimilar insulators under Earth ambient conditions [2]. Moreover, terrestrial dust devils have been observed to emit non-thermal microwave radiation [41], which is presumably produced by micro-discharges between colliding particles [42]. Note, however, that experiments with toner particles argue against the concept of micro-discharges limiting the collisional charge transfer between insulators [43].

Finally, larger scale discharges have also been observed to occur in granular systems of identical insulators [44,45] and could also limit the particle charge.

### C. Semi-empirical adaptation of the model

To allow more flexibility in modeling experiments, a semi-empirical version of the model can be used, in which an effective length scale $\delta_{\text{eff}} = C_E \delta_0$ is used in place of $\delta_0$ in Eqs. (12-16). The empirical parameter $C_E$ thus relates the effective distance from the point of contact over which charge is exchanged ($\delta_{\text{eff}}$) to the theoretical tunneling distance ($\delta_0$), and is introduced to account for processes that cause the charge transfer for non-idealized particles to differ from our purely theoretical considerations. For example, while our simple model assumes perfectly spherical particles and thus a single point of contact during collisions, natural particles are irregular and will contact each other at many separate locations. Moreover, the transfer efficiency of high-energy electrons within the tunneling distance $\delta_0$ will not be unity, because of the energetic constraints that limit the transition of high-energy electrons to empty low-energy states [2,8]. Furthermore, rubbing that may occur during collisions can increase the distance from the 'contact point' for which charge is exchanged beyond the theoretical tunneling distance of Eq. (2) [3]. The empirical parameter $C_E$ can also account for the fact that the transfer efficiency of high-energy electrons is a function that depends on distance, rather than being a step function at the distance $\delta_0$. Note that $C_E$ strongly affects the time constant $\tau = \sum_j N_j \tau_j / \sum_j N_j$ (see Eq. 14) with which charging takes place, but does not affect the final charges for large $t/\tau$. Measurements of the characteristic charging time of granular systems of identical insulators could thus determine $C_E$ for a particular material [39] without any knowledge of the density of trapped high-energy electrons.

### V. SUMMARY AND CONCLUSIONS

We show that the widely observed size-dependent triboelectric charging of chemically identical insulators [9-15,35] is partially due to simple geometrical considerations that produce a net transfer of electrons from larger to smaller particles. This charging mechanism supplements the previously identified 'multiple collisions' charging mechanism [26,31]. We combined these two related mechanisms into the first quantitative scheme of the size-dependent charging of a granular system of chemically identical insulators. Based solely on theoretical considerations, predictions of our charging scheme are qualitatively and quantitatively consistent with measurements of both the magnitude and the characteristic time scale of the charging. Our theory thus seems to provide an explanation for the hitherto puzzling phenomenon of the size-dependent charging of granular systems of identical insulators.

Our charging scheme can be used to study the electrification of a wide range of granular systems, including fluidized beds [9,36-38], powder handling [10], wind-blown sand and snow [11-14,17,19], dust storms and dust devils [18,20], thunderstorms [8,21], and volcanic eruptions [15].

Careful measurements are required to further illuminate the basic physical processes underlying the triboelectric charging of identical insulators and to test and refine our charging scheme.

## ACKNOWLEDGEMENTS

We thank Shanna Shaked for comments on the manuscript. This research was supported by the University of Michigan Rackham Graduate School and by National Science Foundation grants DMR-0705191 and ATM 0622539.

## APPENDIX: CHARGE TRANSFER BETWEEN CUBICAL PARTICLES

We showed in Section IIIa that, during collisions between spherical particles, more trapped high-energy electrons transfer from the larger to the smaller particle than vice versa. In this appendix, we show that a similar effect occurs for cubical particles.

In addition to the assumptions described in Section II, we assume that the difference in the size of the colliding cubical particles is large, such that the collision is synonomous to that of a cubical particle colliding with a flat plane (Fig. 5). Moreover, for simplicity we neglect variations in the angle of rotation perpendicular to the plane that is illustrated in Fig. 5. Accounting for variations in this angle significantly complicates the below derivation, while it yields the same qualitative result.

From the illustration of the charge transfer in Fig. 5a, we find that the number of electrons transferred from the high-energy state of the smaller particle to empty low-energy states on the larger particle equals

$$-\Delta n_{S,H} = -\Delta n_{L,L} = \rho_{H,0} \delta_0 \left( \frac{D_S}{\sin \gamma} + \frac{D_S}{\cos \gamma} + \frac{\delta_0}{\sin \gamma \cos \gamma} \right), \tag{A1}$$

where the subscripts $S$ and $L$ respectively refer to the smaller and larger colliding particle, $D_S$ is the radius of the smaller particle, and the angle $\gamma$ is defined in Fig. 5a. Conversely, the number of trapped high-energy electrons transferred from the larger to the smaller particle equals (see Fig. 5b)

$$-\Delta n_{L,H} = \Delta n_{S,L} = \rho_{H,0} \delta_0 \left[ \frac{D_S}{\sin \gamma} + \frac{D_S}{\cos \gamma} + \frac{\pi \delta_0}{2} \left( \frac{1}{\sin \gamma} + \frac{1}{\cos \gamma} \right) \right]. \tag{A2}$$

As was the case with spherical particles, the charge transfer between cubical particles of different sizes is thus asymmetric. This asymmetry is due to the unequal contributions near the side of the small cube (that is, the shaded triangle in Fig. 5a for the smaller particle and the ovals in Fig. 5b for the larger particle). By subtracting Eqs. (A1) and (A2), we find that the net charge transfer experienced by the larger particle equals

$$\Delta q_L = \frac{e \rho_{H,0} \delta_0^2}{\sin \gamma \cos \gamma} \left[ \frac{\pi}{2} (\sin \gamma + \cos \gamma) - 1 \right]. \tag{A3}$$

The above equations are valid for $|\sin\gamma| \geq \delta_0/D_S$. Since $D_S \gg \delta_0$ we can safely neglect the small contribution from $|\sin\gamma| < \delta_0/D_S$. We thus obtain the average charge transfer between colliding particles by integrating over the angle $\gamma$

$$\overline{\Delta q_L} \approx \frac{4}{\pi} e\rho_{H,0}\delta_0^2 \int_{\arcsin\frac{\delta_0}{D_S}}^{\pi/4} \frac{\frac{\pi}{2}(\sin\gamma+\cos\gamma)-1}{\sin\gamma\cos\gamma}d\gamma \approx 2e\rho_{H,0}\delta_0^2\left[\left(1-\frac{2}{\pi}\right)\ln\frac{D_S}{\delta_0}+\ln 2\right], \quad (A4)$$

where we again used that $D_S \gg \delta_0$ and that $\cos(\arcsin x) = \sqrt{1-x^2}$. In comparison with Eq. (10), the charge transfer between cubical particles is approximately two to four times as large as that between spherical particles.

FIGURES

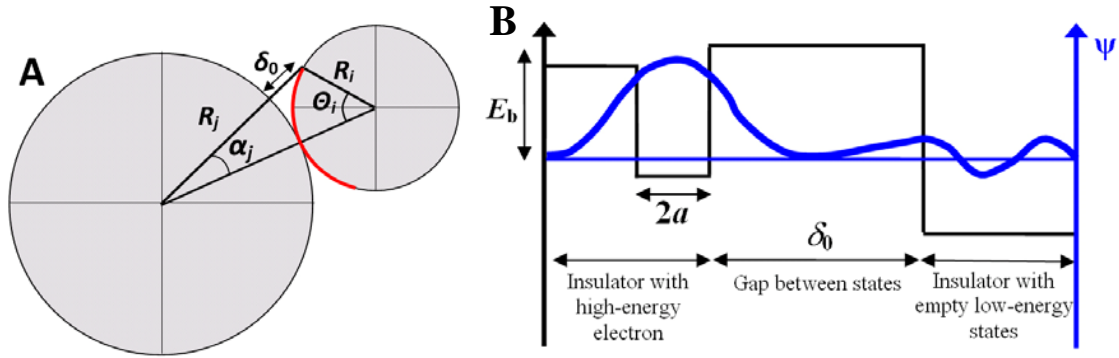

FIG. 1 (color online). (a) Schematic of the charge transfer occurring during a collision between two spherical particles of identical material but of different sizes $R_i$ and $R_j$. The angle $\theta_i$ represents the maximum angle from the point of contact on particles $i$ and $j$ from which trapped high-energy electrons can transfer to empty low-energy states on the opposite particle. The area in which electrons can transfer in this manner is indicated by the thick red arc. (b) Simplified schematic representation of the wavefunction of a high-energy electron in an electron trap near the surface of another insulator with empty low-energy states. The thin black line denotes the electron's potential energy as a function of position, and the thick blue line denotes its wavefunction. After Fig. 1 in Lowell [32].

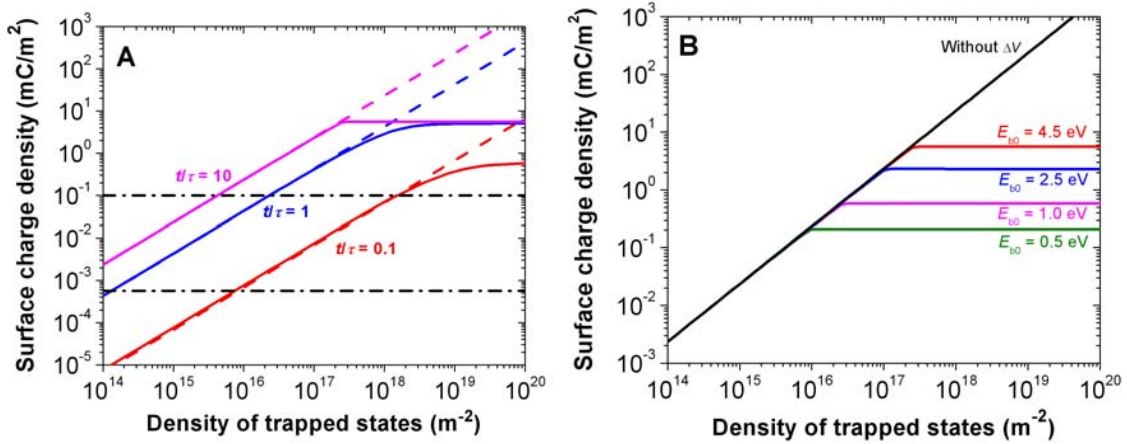

FIG. 2 (color online). **(a)** Average absolute value of the surface charge density as a function of the density of trapped electron surface states for a binary mixture of two particle sizes for $t/\tau = 0.1$, 1, and 10. Dashed and solid lines respectively denote predictions of the analytical (Eq. 16) and numerical (Eq. 11) solutions. The results in this figure depend on the ratio of particle sizes, which we took as 1:4, and on the barrier energy (see Fig. 1b), which we took as $E_{b0} = 4.5$ eV [32]. Note that the results are independent of all other parameters, such as the particle concentration and relative velocity. Those parameters only affect the characteristic charging time $\tau$ (see Eq. 14). The dash-dotted lines represent lower and upper limits on particle charge densities measured in wind-blown sand and dust devils [11,35].
**(b)** Same as part **(a)**, except for different values of the barrier energy $E_{b0}$, with $t/\tau = 10$.

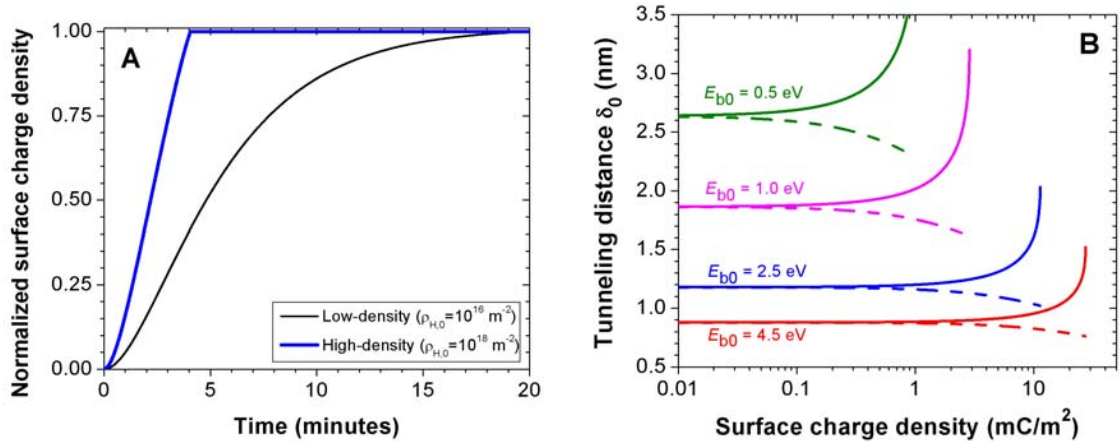

FIG. 3 (color online). **(a)** Normalized surface charge density as a function of time for the low ($\rho_{H,0} = 10^{16}$ m$^{-2}$; thin black line), and high ($\rho_{H,0} = 10^{18}$ m$^{-2}$; thick blue line) limits of the density of trapped electron surface states for $E_{b0} = 4.5$ eV. We use parameters typical for fluidized beds [11], with particle sizes of 100 and 400 μm which both occupy 25% of the volume, and a relative velocity between the particles of 0.1 m/s. The average absolute surface charge density is normalized to its value at 20 minutes. We find a characteristic charging time $\tau$ of several minutes, which agrees well with measurements of fluidized beds [36-38].

**(b)** The tunneling distance $\delta_0$ (see Eqs. 2 – 5) for two particles with equal but opposite surface charge density. The solid (dashed) lines denote $\delta_0$ for electrons tunneling from the negatively (positively) charged particle to the positively (negatively) charged particle. The predicted tunneling distance is on the order of 1-2 nm, which is consistent with previous literature estimates [1-3,32].

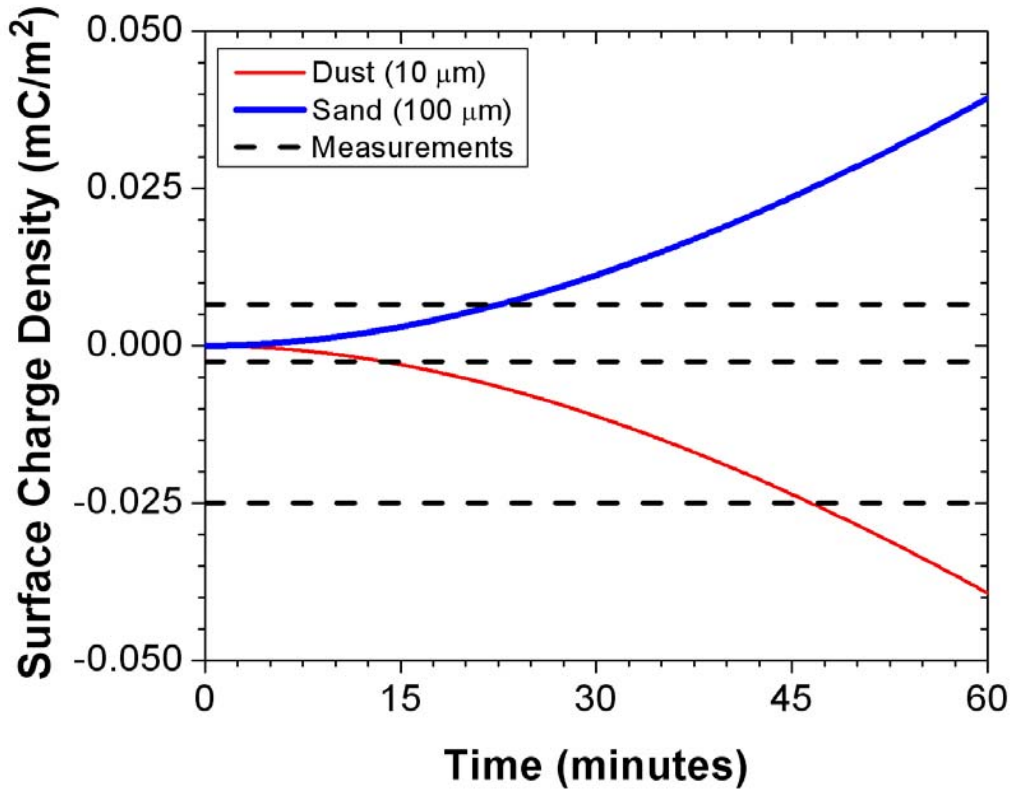

FIG. 4 (color online). Application of our charging scheme to dust storms and dust devils. Charging occurs due to collisions between dust (~10 μm; thin red line) and sand (~100 μm; thick blue line) [17,20]. We used $\rho_{H,0} = 10^{16}$ m$^{-2}$ and $E_{b0} = 4.5$ eV, assumed a relative velocity of 1 m/s, used mass loadings of 10 g/m$^3$ for the dust [46] and 100 g/m$^3$ for the sand [17], and used a particle density of 2650 kg/m$^3$ [17]. The predicted characteristic charging time is approximately two hours. The top dashed line indicates measurements of the average charge density of saltating sand [11], while the bottom two dashed lines denote lower and upper limits of measurements of the negative charge density held by individual 10 μm dust particles [35]. Note that the numerical (Eq. 16) and analytical (Eq. 11) solutions, which respectively do and do not account for the effect of particle charges on subsequent charge transfer during collisions, yield identical results (see also Figure 2a).

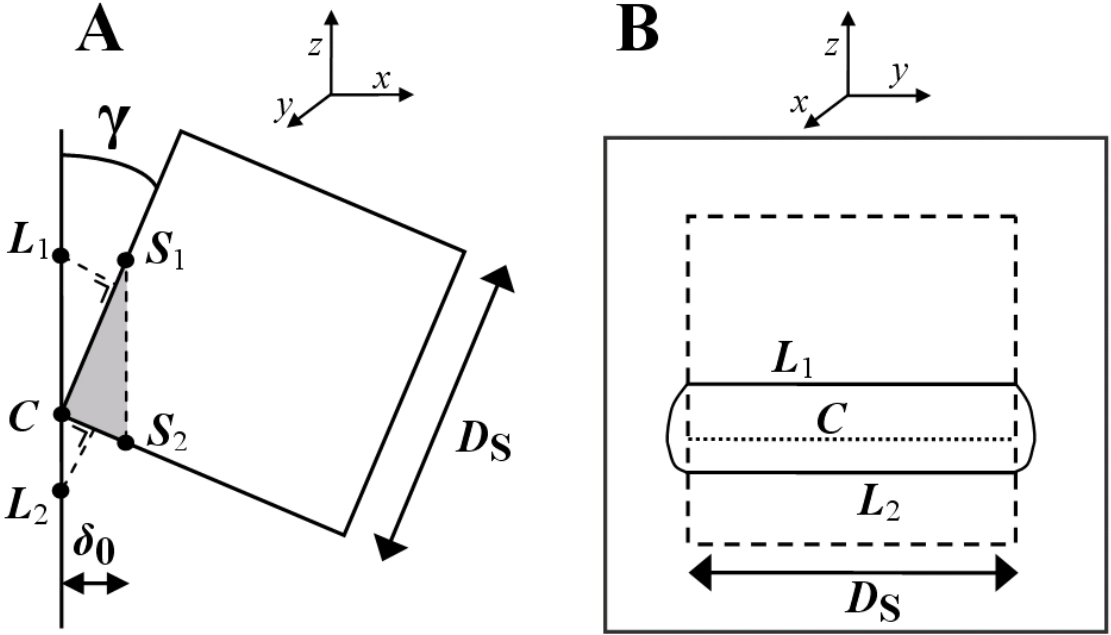

Fig. 5. (**a**) Schematic in the *xz*-plane of a collision between a small cubical particle with diameter $D_S$ and a much larger cubical particle with diameter $D_L$. The distance of the points $S_1$, $S_2$, $L_1$, and $L_2$ to the opposite particle is exactly $\delta_0$. These points respectively lie at $(\delta_0, \delta_0/\tan\gamma)$, $(\delta_0, -\delta_0\tan\gamma)$, $(0, \delta_0/\sin\gamma)$, and $(0, -\delta_0/\cos\gamma)$, relative to the point of contact $C$ in the *xz*-plane. The total area of the smaller cube that lies within a distance $\delta_0$ of the larger cube is thus $A = D_S \Delta CS_1 + D_S \Delta CS_2 + \Delta CS_1 \Delta CS_2$ (see Eq. A1), where $\Delta CS_1 = \delta_0/\sin\gamma$ and $\Delta CS_2 = \delta_0/\cos\gamma$ are the distances of $S_1$ and $S_2$ from $C$, and the shaded triangle denotes the area $0.5 \Delta CS_1 \Delta CS_2$.

(**b**) As in (**a**), except for the *yz*-plane. The outer gray square represents the larger cube, and the dashed rectangle denotes the projection of the smaller cube on the surface of the larger cube. The dotted line represents the line of contact between the particles, and the solid line denotes the points on the larger cube for which the distance to the smaller cube is exactly equal to $\delta_0$. The two semi-ovals are described by the equations $y = \pm\sqrt{\delta_0^2 - z^2 \sin^2\gamma}$ ($z > 0$) and $y = \pm\sqrt{\delta_0^2 - z^2 \cos^2\gamma}$ ($z < 0$). The total area on the larger cube over which high-energy electrons can tunnel to the smaller cube is thus obtained by integrating these equations (which produces the third term in the brackets on the right-hand side in Eq. A2) and adding the result to the area between the lines $L_1$ and $L_2$ (which produces the first two terms in Eq. A2).